\documentclass{article}
\usepackage{spconf,amsmath,graphicx, url, tikz,pgfplots,multirow, booktabs,bm}
\usepackage{atbegshi,picture}

\newcommand{\etal}{\textit{et al.}}
\newcommand{\dB}{\,dB~}
\title{SPATIALLY-ADAPTIVE LEARNING-BASED IMAGE COMPRESSION WITH HIERARCHICAL MULTI-SCALE LATENT SPACES}
\name{Fabian Brand, Alexander Kopte, Kristian Fischer, André Kaup}
\address{Multimedia Communications and Signal Processing\\ Friedrich-Alexander-Universit\"at Erlangen-N\"urnberg\\
	Cauerstr. 7, 91058 Erlangen}
\AtBeginShipout{\AtBeginShipoutUpperLeft{%
		\put(\dimexpr\paperwidth-20cm\relax,-1.5cm){\parbox[t]{19cm}{\copyright 2023 IEEE.  Personal use of this material is permitted.  Permission from IEEE must be obtained for all other uses, in any current or future media, including reprinting/republishing this material for advertising or promotional purposes, creating new collective works, for resale or redistribution to servers or lists, or reuse of any copyrighted component of this work in other works.}}%
}}

\begin{document}
\ninept
\maketitle

\makeatletter
\def\ps@IEEEtitlepagestyle{%
	\def\@oddfoot{\mycopyrightnotice}%
	\def\@oddhead{\hbox{}\@IEEEheaderstyle\leftmark\hfil\thepage}\relax
	\def\@evenhead{\@IEEEheaderstyle\thepage\hfil\leftmark\hbox{}}\relax
	\def\@evenfoot{}%
}
\def\mycopyrightnotice{%
	\begin{minipage}{\textwidth}
		\scriptsize
		\copyright 2023 IEEE.  Personal use of this material is permitted. Permission from
		IEEE must be obtained for all other uses, in any current or future
		media, including reprinting/republishing this material for advertising
		or promotional purposes, creating new collective works, for resale or
		redistribution to servers or lists, or reuse of any copyrighted
		component of this work in other works.
	\end{minipage}
}
\makeatother

\begin{abstract}
	\vspace{-0.05cm}
Adaptive block partitioning is responsible for large gains in current image and video compression systems. This method is able to compress large stationary image areas with only a few symbols, while maintaining a high level of quality in more detailed areas. Current state-of-the-art neural-network-based image compression systems however use only one scale to transmit the latent space. In previous publications, we proposed RDONet, a scheme to transmit the latent space in multiple spatial resolutions. Following this principle, we extend a state-of-the-art compression network by a second hierarchical latent-space level to enable multi-scale processing. We extend the existing rate variability capabilities of RDONet by a gain unit. With that we are able to outperform an equivalent traditional autoencoder by 7\% rate savings. Furthermore, we show that even though we add an additional latent space, the complexity only increases marginally and the decoding time can potentially even be decreased. 
\end{abstract}
\vspace{-0.15cm}
\vspace{-0.2cm}
\section{Introduction}
\label{sec:intro}

With the rise of neural networks for image processing, deep-learning-based image compression methods are widely researched topics. A large portion of networks are based on the autoencoder architecture, where an image is transformed into a low-dimensional latent space, which is then transmitted over the channel, before the image is reconstructed from the retrieved latent representation. To efficiently transmit the latent representation, the coder also estimates and transmits a probability model.

Currently these systems are able to outperform classical methods like VVC~\cite{BrossWY2021_OverviewVersatileVideo} intra coding across all distortion measures~\cite{ChengST2020_LearnedImageCompression,KoyuncuGB2022_Contextformertransformerspatio}. These gains are largely due to two developments: First, the encoder and decoder functions got more and more sophisticated over time. While first proposals only used three or four convolutional layers~\cite{BalleLS2017_Endendoptimized,BalleMS2018_Variationalimagecompression,MinnenBT2018_JointAutoregressiveHierarchical}, current architectures contain residual layers, attention mechanisms~\cite{ChengST2020_LearnedImageCompression}, or even transformers~\cite{ZouSZ2022_DevilIsDetails}. This influences the quality of the features which are transmitted, finding more compact and decorrelated representations. A second line of improvements was made with the probability modeling. Early on, the probability model was improved from fixed models for the entire latent space~\cite{BalleLS2017_Endendoptimized}, over parametrized models where the parameter is transmitted via a side channel~\cite{BalleMS2018_Variationalimagecompression} to an autoregressive model~\cite{MinnenBT2018_JointAutoregressiveHierarchical}, which makes use of already transmitted elements to model the probabilities for the next ones. This model was subsequently modified from a spatial autoregressive model to a channel autoregressive model~\cite{MinnenS2020_ChannelWiseAutoregressive} or to combined models~\cite{HeYP2022_ELICEfficientLearned,KoyuncuGB2022_Contextformertransformerspatio}. Also the probability model itself was extended from a Gaussian model in~\cite{MinnenBT2018_JointAutoregressiveHierarchical} to the use of Gaussian mixture models (GMMs) in~\cite{ChengST2020_LearnedImageCompression}. In the first papers, one model had to be trained for each rate point, which is in strong contrast to traditional methods, where the same coding software can produce multiple rate points. To solve this problem, Cui \etal proposed~\cite{CuiWG2021_AsymmetricGainedDeep} the use of gain units which adaptive scale the latent space depending on the desired rate-distortion trade-off.

A source of large gains in traditional image and video compression is the processing in different scales dependent on the image content. This concept was successfully applied for example in HEVC~\cite{SullivanOH2012_OverviewHighEfficiency}, VVC~\cite{BrossWY2021_OverviewVersatileVideo} and, AV1~\cite{HanLM2021_TechnicalOverviewAV1}. Here, the coder uses larger blocks to compress areas with little structure and finer blocks for detailed areas. Among others, this has the advantage that stationary areas can be transmitted in large chunks by only signaling few coefficients. On the other hand, deep-learning-based compression has mainly remained at a $16\!\times\!16$ downsampling which was used in the pioneering paper~\cite{BalleLS2017_Endendoptimized}. We propose a novel compression network, which is based on previous works~\cite{BrandFK2021_RateDistortionOptimized,BrandFK2022_LearningTrueRate_DCC,BrandFK2022_RDONetRateDistortion}. We extend this previous work to a stronger, fully functional image coder including attention mechanisms and gain units. This two latent space approach shows significant improvements over an equivalent standard autoencoder, which has similar performance as VTM Intra. Compared to our previous work, this is the first time, that we assess the performance of the multi-scale network RDONet on objective metrics using a state-of-the-art network. We demonstrate that RDONet can increase the quality range covered by the gain-units, which is a large advantage for practical applications. Even though we use two latent spaces, we show that the overall complexity is comparable to the baseline and the decoding time can even decrease.
\vspace{-0.2cm}
\section{Related Work}
Most neural-network-based image compression methods use fixed transforms which are not changed once the training is done. In traditional compression, it has been shown that content-adaptive encoding and decoding functions can yield large gains. This is for example the case in adaptive block partitioning or in signaled intra prediction modes, where structured side-information is used to improve the compression performance. This is typically done using a rate-distortion optimization during encoding. There are some examples, where rate-distortion optimization is used in learning-based compression. In~\cite{RozendaalHC2021_OverfittingFunProfit}, the authors modify the weights of the decoder and transmit the model updates, which showed large gains over the baseline. In~\cite{SchaferPP2021_RateDistortionOptimization}, the authors propose a fast search scheme to optimize the quantized latent representation which is transmitted over the channel. In~\cite{BrandFK2021_RateDistortionOptimized}, we proposed the first image compression network, which implements adaptive transmission in different hierarchical latent spaces with decreasing resolution. This method is conceptually similar to adaptive block partitioning.

\section{Spatially Adaptive Image Compression}\vspace{-0.2cm}
In this work, we propose a compression system which can compress the latent representations on two scales. One high resolution scale, which transmits the features after the typical downsampling with a factor of 16 in each direction, and one low resolution scale, which increases the downsampling factor of the image to 32. The decision, which scale to use can be made at encoding time and is transmitted on a side-channel.

To achieve such a multi-scale transmission, the network structure has to be modified. The presented method is modular and can be applied on any autoencoder-based compression system. To construct the changes in the network, we split the network in a feature generating function $f_a(\bm{x})$, to generate the latent representation $\bm{y}$ from the image $\bm{x}$, an image synthesis function $f_s(\hat{\bm{y}})$, which reconstructs the image $\hat{\bm{x}}$ from the quantized latent space $\hat{\bm{y}}$, and a transmission block, which transmits the latent representation, which we call latent space unit (LS-Unit). The LS-Unit contains quantization, context model, hyperprior and arithmetic encoder and decoder.

\begin{figure}[t!]
	\centering
	\includegraphics[width=0.75\linewidth]{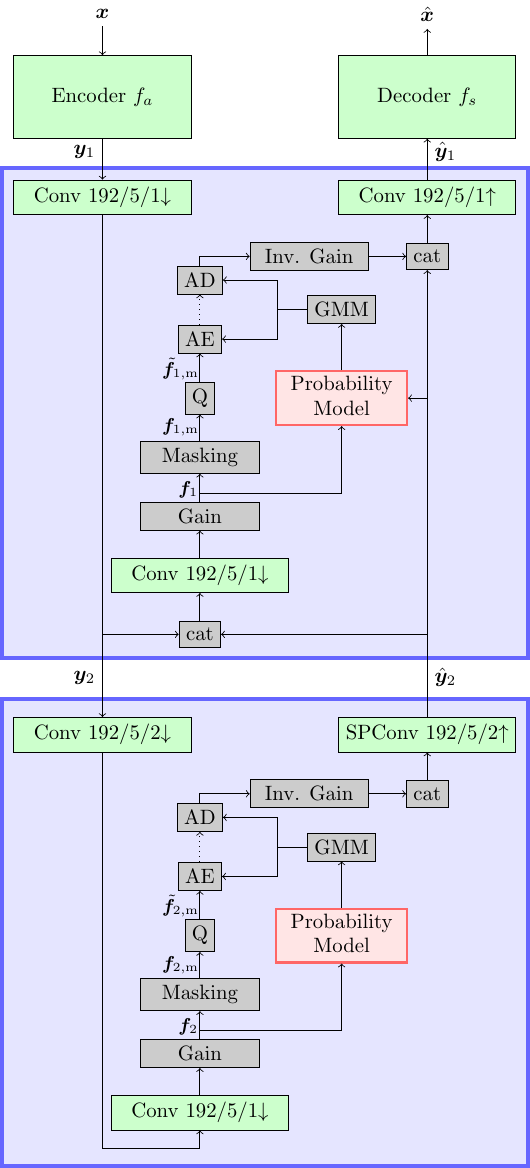}
	\caption{Schematic of RDONet with two latent space units (LS-Units). Transmission over the channel is performed within the LS-Unit. The LS-Units are marked with a blue background. The probability model contains a context model and a hyperprior network, which is a conditional autoencoder in the first latent space unit. SP-Conv denotes subpixel convolutional layers as in~\cite{ShiCH2016_RealTimeSingle}\label{Fig:Schematic}\vspace{-0.2cm}}
\end{figure}

The structure of the proposed network is shown in Fig.~\ref{Fig:Schematic}. We see that the network contains two LS-Units instead of one. Note that even though it is not called latent-space unit, standard image coders contain exactly one LS-Unit, which is usually simpler and directly transmits the latent representation. To achieve compression at multiple scales, we need to apply further downsampling to the latent representation within the second LS-Unit to transmit a lower dimensional representation. Note that we keep the number of channels constant, so that with each downsampling step, the coder is forced to further compress the information. To enable multi-scale processing and passing on representations to lower LS-Units, we also need to make some changes to the LS-Unit itself. Most importantly, we need to add a masking block, which, based on external input, masks out individual positions of the latent representation by setting them to zero. These elements are not transmitted, since the decoder knows their value via the mask. That way it is possible to compress each part of the image in individual spatial resolution. Further, we connect the latent spaces to exploit that the different representations have redundancy between them. We therefore state that the latent spaces are transmitted in descending order, from the lower-dimensional latent space to the higher. We can condition the LS-Units on the previously decoded representation $\hat{\bm{y}}_2$. We do this in two ways, once by combining the current representation $\bm{y}_2$ with the previously decoded representation $\hat{\bm{y}}_2$ via a convolution. This way, we obtain a kind of generalized difference, which we also successfully employed for conditional P-frame coding in~\cite{BrandSK2022_PFrameCoding}. Furthermore, we use a conditional hyperprior such that the previously known information can be taken into account within the probability model. This additional convolutional layer also takes into account that the representation $\bm{y}_2$, which is passed on, may contain features which are only relevant for the next latent space and hence should not be present in the transmission of the current latent space.

In previous work on RDONet, we were able to obtain multiple rate-points per trained model by assigning more or less areas to the lower dimensional latent space. However, we also showed that this works only in a small range around the optimal point, and that quality deteriorates if the distance to the optimal rate-point becomes too large. In this work, we want to improve this behavior by including gain units similar as~\cite{CuiWG2021_AsymmetricGainedDeep} to support a wider range of multiple rate points. In preliminary experiments, we found that sampling $\lambda$ only from the supporting points can sometimes lead to unstable behavior, since the coder is not used to gain interpolation during the training. We therefore randomly picked an interval between two values and then selected a random value within this interval. 

\section{Experiments and Results}
\subsection{Network Structure and Training}
For the experiments in this work, we use the network structure by Cheng~\etal~\cite{ChengST2020_LearnedImageCompression} as backbone. We therefore use the encoding and decoding network exactly as in their work and take over the main components of the probability model, in particular, we use a GMM with three Gaussians and choose the network to estimate the GMM parameters in the same way. The only modification was to use a small context model also to transmit the hyperprior, as we did in our previous publications~\cite{BrandFK2022_LearningTrueRate_DCC}.

For the baseline, we trained this model on 80000 pictures from the Unsplash dataset\footnote{\url{https://unsplash.com/data}} for a total of 1 million iterations, as suggested in~\cite{ChengST2020_LearnedImageCompression}. We used an Adam~\cite{KingmaB2015_AdamMethodStochastic} optimizer with an initial learning rate of 10$^{-4}$, which we decreased to 10$^{-5}$ for the last 80000 iterations. During the training, we sampled $\lambda$ randomly from the intervals for the supporting points \mbox{$\Lambda = [0.000625, 0.0025, 0.01, 0.02, 0.04, 0.08]$}.

We trained the network on the rate distortion loss function
\begin{equation}
\mathcal{L}=\mathcal{D}+\lambda\mathcal{R},
\end{equation} 
where $\mathcal{R}$ is the estimated rate to transmit all latent spaces. As a distortion $\mathcal{D}$, we use the mean squared error (MSE) and a weighted combination of MSE and the multiscale structural similarity index measure (MS-SSIM)~\cite{WangSB_Multiscalestructuralsimilarity}. The first choice of loss will result in a model which aims to reconstruct the image as closely as possible, while the second one will yield more perceptually pleasant images. We choose a combination of both losses since preliminary experiments have shown that training only on MS-SSIM can produce unstable behavior in the early convergence. We weight the MS-SSIM with a factor of 0.1 before adding it to the MSE. Note that due to the different order of magnitudes, MS-SSIM is still the dominant distortion loss. When training on MSE, we weight the MSE with a factor of 5 in order to achieve similar rate points with the same set $\Lambda$.

When training RDONet, we used a similar strategy as we proposed in~\cite{BrandFK2022_LearningTrueRate_DCC}, where we first chose random masks during the first 400,000 iterations and then use a variance threshold criterion, where all areas with a pixel-value variance of 0.002 or lower are coded in the low resolution latent space.

\subsection{Rate Distortion Performance}
We evaluate the network on the TECNICK~\cite{AsuniG2013_TESTIMAGESLargeData} dataset. In all our experiments, we generate a decodable bitstream and measure the filesize to obtain the rate. This bitstream also contains the information about the mask. To measure the distortion, we use PSNR and MS-SSIM~\cite{WangSB_Multiscalestructuralsimilarity}, as implemented in the \texttt{pytorch\_msssim} library. As commonly done, we report MS-SSIM in dB using the following formula
\begin{equation}
	\text{MS-SSIM}_\mathrm{dB} = -10\log_{10}(1-\text{MS-SSIM})
\end{equation}
We compare both methods against VTM~\cite{BrossWY2021_OverviewVersatileVideo} version 18.2 in 4:4:4 chroma configuration wit h10 bits. Otherwise, we kept the standard parameters of VTM in all intra configuration. We use QPs from 30 to 42 to match the rate points obtained with our networks.

Note that RDONet offers in principle the possibility to perform rate-distortion-optimization to obtain optimal depth levels for all parts of the image. This search is computationally very expensive. To have a fair comparison with the baseline in terms of encoding complexity, we use RDONet with the variance-based zero-pass RDO, which we proposed in~\cite{BrandFK2022_LearningTrueRate_DCC}.

\begin{figure}[t]
	\begin{tabular}{c}
		\includegraphics[]{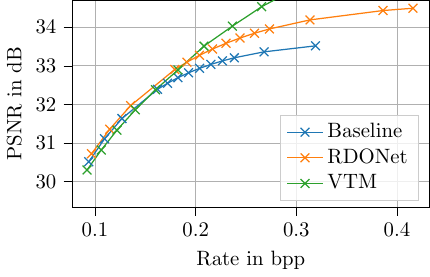}\\
	\includegraphics[]{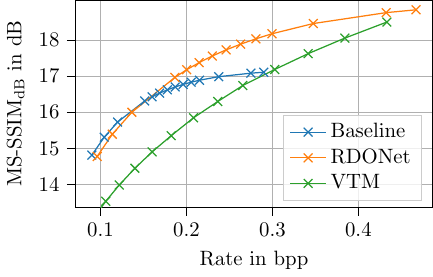}
	\end{tabular}
\vspace{-0.4cm}
\caption{Rate-distortion curves for the TECNICK dataset of the non-hierarchical baseline, RDONet and VVEnc. For the top plot, which shows PSNR, the networks were trained on MSE, for the bottom plot, which shows MS-SSIM, the networks were trained on the combined loss.\label{Fig:Result}\vspace{-0.6cm}}
\end{figure} 
\begin{table}[t]
	\begin{tabular}{c|c|rrr}
		\toprule
		Anchor.            &          Metric          &  Coder   & BD Rate  & BD Quality \\ \midrule
		\multirow{4}{*}{VVEnc}   &  \multirow{2}{*}{PSNR}   & Baseline & +3.11\% & -0.38\dB  \\
		&                          &  RDONet  & +2.29\% & -0.34\dB   \\
		& \multirow{2}{*}{MS-SSIM} & Baseline & -36.13\% &  +1.30\dB \\
		&                          &  RDONet  & -33.78\% &  +1.29\dB  \\ \midrule
		\multirow{2}{*}{Baseline} &           PSNR           &  RDONet  & -7.35\% &  +0.29\dB   \\
		&         MS-SSIM          &  RDONet  &  -1.06\% &  +0.21\dB  \\ \bottomrule
	\end{tabular}
	\vspace{-0.2cm}
	\caption{BD rate and quality. We show the gains of both network-based method over VTM and the gain of RDONet over the baseline. We report the gains with the PSNR and the MS-SSIM metric.\label{Tab:Results}\vspace{-0.5cm}}
\end{table}

We show the results in Fig.~\ref{Fig:Result}. We see that both network-based methods outperform VTM for smaller rates on PSNR and loose for higher rates when they reach a saturation. However, RDONet saturates slower and on a significantly higher quality level. The better performance in higher rates can be explained with the presence of a lower resolution latent space. Some areas, in particular areas with very little texture, can be coded with only very little rate also for high qualities. RDONet offers this possibility by transmitting them in the lower latent space. The higher latent space can therefore transmit with a higher rate yielding a better quality in complex image areas. We furthermore see that the hierarchical structure of RDONet allows for a better flexibility and covers a wider rate-range than the baseline. As expected since the network were also optimzed on MS-SSIM, both neural network-based coders outperform VTM clearly on MS-SSIM. Here, the non-hierarchical baseline performs slightly better in the lower rates, however, RDONet clearly performs better in the high rates, being able to keep the quality without saturation for a large range of rates

We also compute Bj\o ntegaard delta~\cite{Bjontegaard2001_CalculationaveragePSNR} (BD) rates for our coders. Since the overlap on the quality axis is sometimes low, we also compute average quality gains (BD quality). Tab.~\ref{Tab:Results} shows the results. On PSNR, RDONet needs 2.29\% more rate than VTM compared to the baseline, which needs 3.11\% more. Note however that the BD rate computation cuts of at different qualities during this computation. The fairest comparison is therefore given in the direct comparison, where we see that RDONet saves 7.35\% rate and gives an average gain of 0.29 dB at the same rate. For MS-SSIM, the BD rate shows larger gains for the non-hierarchical baseline as for RDONet. This seems to contradict the RD curves, however, this is also caused by different cut-off values. In the direct comparison between the baseline and RDONet, we see that on average, RDONet can save 1.06\% rate at the same quality and can increase the MS-SSIM by 0.21 dB at the same rate
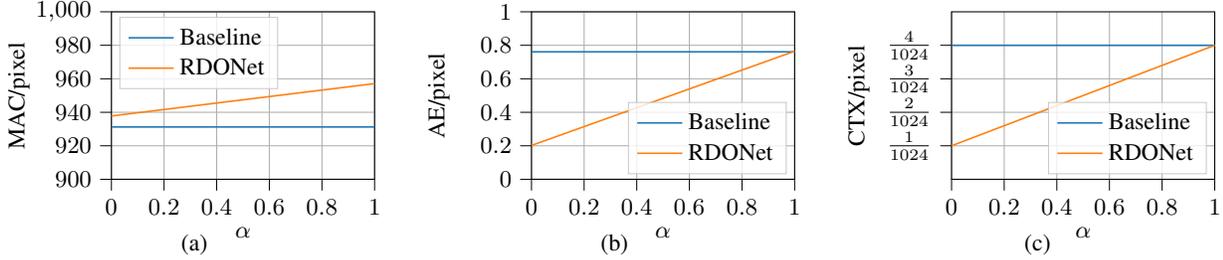
\begin{figure*}
	\begin{tabular}{ccc}
		% This file was created with tikzplotlib v0.10.1.
\begin{tikzpicture}

\definecolor{crimson2143940}{RGB}{214,39,40}
\definecolor{darkgray176}{RGB}{176,176,176}
\definecolor{darkorange25512714}{RGB}{255,127,14}
\definecolor{forestgreen4416044}{RGB}{44,160,44}
\definecolor{lightgray204}{RGB}{204,204,204}
\definecolor{steelblue31119180}{RGB}{31,119,180}

\pgfplotsset{scaled x ticks=false}
\begin{axis}[
width=2in,
height=1.5in,
legend cell align={left},
legend style={
  fill opacity=0.8,
  draw opacity=1,
  text opacity=1,
  at={(0.03,0.97)},
  anchor=north west,
  draw=lightgray204
},
tick align=outside,
tick pos=left,
x grid style={darkgray176},
xlabel={$\alpha$},
xmin=0, xmax=1,
xtick style={color=black},
y grid style={darkgray176},
ylabel={MAC/pixel},
ymin=900, ymax=1000,
ytick style={color=black},
grid = major
]
\addplot [semithick, steelblue31119180]
table {%
0 931.240
1 931.240
};
\addlegendentry{Baseline}
\addplot [semithick, darkorange25512714]
table {%
	0 937.794
	1 957.125
};
\addlegendentry{RDONet}
\end{axis}

\end{tikzpicture}&% This file was created with tikzplotlib v0.10.1.
\begin{tikzpicture}

\definecolor{crimson2143940}{RGB}{214,39,40}
\definecolor{darkgray176}{RGB}{176,176,176}
\definecolor{darkorange25512714}{RGB}{255,127,14}
\definecolor{forestgreen4416044}{RGB}{44,160,44}
\definecolor{lightgray204}{RGB}{204,204,204}
\definecolor{steelblue31119180}{RGB}{31,119,180}

\pgfplotsset{scaled x ticks=false}
\begin{axis}[
width=2in,
height=1.5in,
legend cell align={left},
legend style={
  fill opacity=0.8,
  draw opacity=1,
  text opacity=1,
  at={(0.97,0.03)},
  anchor=south east,
  draw=lightgray204
},
tick align=outside,
tick pos=left,
x grid style={darkgray176},
xlabel={$\alpha$},
xmin=0, xmax=1,
xtick style={color=black},
y grid style={darkgray176},
ylabel={AE/pixel},
ymin=0, ymax=1,
ytick style={color=black},
grid = major
]
\addplot [semithick, steelblue31119180]
table {%
0 0.761
1 0.761
};
\addlegendentry{Baseline}
\addplot [semithick, darkorange25512714]
table {%
	0 0.2021
	1 0.7646
};
\addlegendentry{RDONet}
\end{axis}

\end{tikzpicture}&% This file was created with tikzplotlib v0.10.1.
\begin{tikzpicture}

\definecolor{crimson2143940}{RGB}{214,39,40}
\definecolor{darkgray176}{RGB}{176,176,176}
\definecolor{darkorange25512714}{RGB}{255,127,14}
\definecolor{forestgreen4416044}{RGB}{44,160,44}
\definecolor{lightgray204}{RGB}{204,204,204}
\definecolor{steelblue31119180}{RGB}{31,119,180}

\pgfplotsset{scaled x ticks=false}
\begin{axis}[
width=2in,
height=1.5in,
legend cell align={left},
legend style={
  fill opacity=0.8,
  draw opacity=1,
  text opacity=1,
  at={(0.97,0.03)},
  anchor=south east,
  draw=lightgray204
},
tick align=outside,
tick pos=left,
x grid style={darkgray176},
xlabel={$\alpha$},
xmin=0, xmax=1,
xtick style={color=black},
y grid style={darkgray176},
ylabel={CTX/pixel},
ymin=0, ymax=5,
ytick={1,2,3,4},
yticklabels={$\frac{1}{1024}$,$\frac{2}{1024}$,$\frac{3}{1024}$,$\frac{4}{1024}$},
ytick style={color=black},
grid = major
]
\addplot [semithick, steelblue31119180]
table {%
0 4
1 4
};
\addlegendentry{Baseline}
\addplot [semithick, darkorange25512714]
table {%
	0 1
	1 4
};
\addlegendentry{RDONet}
\end{axis}

\end{tikzpicture}\vspace{-0.3cm}\\\vspace{-0.3cm}
		(a) & (b) & (c)
	\end{tabular}
	\caption{Complexity of the baseline network and RDONet. We show MAC/pixel in (a), calls to the arithmetic coder (AE) per pixel in (b) and decoder calls to the context model (CTX) per pixel in (c). $\alpha$ denotes the fraction of the image which is coded in the higher latent space.\label{Fig:Complexity}\vspace{-0.5cm}}
\end{figure*}
\subsection{Complexity}
Even though RDONet adds additional layers for a second latent space, the complexity does not increase much. Depending on how the complexity is measured, it can even decrease the complexity by performing certain time-consuming operations in a lower resolution. Also, the additional complexity of RDONet is only added after four downsampling steps, such that additional operations are relatively light-weight compared to the complex operations, like attention mechanisms in the higher layers.

In the following, we consider three different complexity measures. We first look at multiply-accumulate operations (MAC) per pixel. This measure tells us how many computations need to be performed in order to transform the image to the latent representation and back and to compute the probability models. Second, we look at the number of symbols which have to be written, i.e. how often the arithmetic coder has to be called. We denote this quantity as ``AE''. Third, we look at how often the context model has to be called during decoding. Since the context model is autoregressive, the probability of one symbol is only available after all previous symbols have been decoded. The context model has to be updated after each decoded hyperpixel, which typically increases the decoding runtime, since the parallel processing capabilities of GPUs can not be used. The number of decoder calls of the context model is denoted as ``CTX''.

For the baseline, all complexity measures are constant for all images. For RDONet on the other hand, they depend on the image, in particularly, which portion of the image is coded in the higher latent space and which is coded in the lower latent space. We define $\alpha$ as the portion of the image which is coded in the higher latent space. In the TECNICK dataset, we observe values of $\alpha$ from 0.32 to 0.98 with an average of 0.68. In general, processing in the lower latent space is cheaper in terms of complexity, since larger parts of the image are coded in just one hyperpixel. 

In Fig.~\ref{Fig:Complexity}, we show the complexity depending on $\alpha$. We see that, as expected, all complexity measures increase with $\alpha$. For the number of multiply-accumulate operations (a), we see that both the best and worst case of RDONet lie above the baseline. However, since the additional layers are all located deep in the network with low spatial dimension, we only see an increase of MAC/pixel by between 0.7\% and 2.8\%. This is mainly due to the overhead of splitting up and putting back together the latent spaces. This could be further optimized, since some of the values are known to be zero from the masking information. This would depend on the exact distribution of the mask, since these layers have a field of view of more than one pixel. Such a connection is difficult to model, therefore we assume the worst case in our evaluation, that the entire image is processed in these layers.

In Fig.~\ref{Fig:Complexity}(b), we see the number of calls to the arithmetic coder per pixel depending on $\alpha$. When $\alpha=0$, we see that this number can be drastically reduced compared to the baseline, since transmitting in the lower latent space requires only one fourth of symbols written to the channel. Note that we have a small overhead since we need to transmit two hyperpriors. This overhead is almost negligible, since the latent space is subsampled by a factor of $8\times8$ before the hyperprior information is transmitted.

We observe a similar effect when we look at the number of decoder calls to the context model. When the entire image is transmitted in the lower latent space, the context model needs to be called on once for 1024 pixels, in the highest latent space and in the baseline, the context model needs to be called onces for only 256 pixels, increasing the number of calls by four. 

The overall complexity of a coder largely depends on the system and the implementation of the individual components. However, we see only a small increment in MAC/pixels using RDONet. On the other hand, we are able to decrease the number of decoder calls to the context model. Since this process is inherently sequential, it largely effects the decoding speed compared to the remainder of the network, which can be parallelized. In most situations, RDONet can therefore offer a faster decoding process.
\vspace{-0.2cm}
\section{Conclusion}\vspace{-0.2cm}
In this paper, we propose a method for neural-network-based compression which uses multiple latent spaces for transmission. By building on our previous work, our method RDONet is able to outperform conventional autoencoders, as for example the method proposed by Cheng~\etal~\cite{ChengST2020_LearnedImageCompression}, which we trained in the same way on the same data. The RDONet method is completely modular and, in principle, can be applied on any autoencoder-based compression network, where we expects similar gains. The gain was largely achieved by a larger flexibility due to the two available latent spaces, which yielded in a larger achievable rate range using gain units and allows better bit-allocation. We furthermore showed that RDONet does not increase the inference complexity much and can even decrease decoding runtime since less time consuming sequential calls of the spatial context model are necessary.

In future work, RDONet can be further optimized on complexity to eliminate overhead. Furthermore, the behavior for large rate points for near lossless coding can be researched. The context model in RDONet is confronted with a sparse latent space, which may lead to suboptimal behavior. Solving this problem can probably increase the performance even further. Also, a full RDO is currently very costly. We have already shown that the zero-pass solution comes close to the results of a full RDO, but a significant speed-up of the search can make the full search and the corresponding performance feasible and can achieve RDONets full potential also in practical cases. However, the already superior performance of RDONet compared to the non-hierarchical baseline shows the importance of multi-scale image compression and that structured side-information can yield large gains not only in traditional but also in neural-network-based image compression.
\bibliographystyle{IEEE}

\begin{thebibliography}{10}
	
	\bibitem{BrossWY2021_OverviewVersatileVideo}
	Benjamin Bross, Ye-Kui Wang, Yan Ye, Shan Liu, Jianle Chen, Gary~J. Sullivan,
	and Jens-Rainer Ohm,
	\newblock ``Overview of the versatile video coding ({VVC}) standard and its
	applications,''
	\newblock {\em {IEEE} Transactions on Circuits and Systems for Video
		Technology}, vol. 31, no. 10, pp. 3736--3764, oct 2021.
	
	\bibitem{BalleLS2017_Endendoptimized}
	Johannes Ball{\'e}, Valero Laparra, and Eero~P Simoncelli,
	\newblock ``End-to-end optimized image compression,''
	\newblock in {\em Proc. International Conference on Learning Representations
		(ICLR)}, Apr 2017, pp. 1 -- 27.
	
	\bibitem{BalleMS2018_Variationalimagecompression}
	Johannes Ballé, David Minnen, Saurabh Singh, Sung~Jin Hwang, and Nick
	Johnston,
	\newblock ``Variational image compression with a scale hyperprior,''
	\newblock in {\em Proc. International Conference on Learning Representations
		({ICLR})}, 2018, pp. 1--47.
	
	\bibitem{MinnenBT2018_JointAutoregressiveHierarchical}
	David Minnen, Johannes Ball\'{e}, and George~D Toderici,
	\newblock ``Joint autoregressive and hierarchical priors for learned image
	compression,''
	\newblock in {\em Advances in Neural Information Processing Systems}, Dec.
	2018, vol.~31, pp. 1--10.
	
	\bibitem{ChengST2020_LearnedImageCompression}
	Zhengxue Cheng, Heming Sun, Masaru Takeuchi, and Jiro Katto,
	\newblock ``Learned image compression with discretized gaussian mixture
	likelihoods and attention modules,''
	\newblock in {\em Proc. {IEEE} Conference on Computer Vision and Pattern
		Recognition ({CVPR})}, 2020.
	
	\bibitem{ZouSZ2022_DevilIsDetails}
	Renjie Zou, Chunfeng Song, and Zhaoxiang Zhang,
	\newblock ``The devil is in the details: Window-based attention for image
	compression,''
	\newblock in {\em Proc. {IEEE}/{CVF} Conference on Computer Vision and Pattern
		Recognition ({CVPR})}, June 2022.
	
	\bibitem{MinnenS2020_ChannelWiseAutoregressive}
	David Minnen and Saurabh Singh,
	\newblock ``Channel-wise autoregressive entropy models for learned image
	compression,''
	\newblock in {\em Proc. {IEEE} International Conference on Image Processing
		({ICIP})}, Oct. 2020.
	
	\bibitem{HeYP2022_ELICEfficientLearned}
	Dailan He, Ziming Yang, Weikun Peng, Rui Ma, Hongwei Qin, and Yan Wang,
	\newblock ``Elic: Efficient learned image compression with unevenly grouped
	space-channel contextual adaptive coding,''
	\newblock in {\em Proceedings of the IEEE/CVF Conference on Computer Vision and
		Pattern Recognition (CVPR)}, June 2022, pp. 5718--5727.
	
	\bibitem{KoyuncuGB2022_Contextformertransformerspatio}
	Burakhan Koyuncu, Han Gao, Atanas Boev, Georgii Gaikov, Elena Alshina, and
	Eckehard Steinbach,
	\newblock ``Contextformer: A transformer with spatio-channel attention for
	context modeling in learned image compression,''
	\newblock in {\em Proc. European Conference on Computer Vision (ECCV)}, Oct.
	2022, pp. 447--463.
	
	\bibitem{CuiWG2021_AsymmetricGainedDeep}
	Ze~Cui, Jing Wang, Shangyin Gao, Tiansheng Guo, Yihui Feng, and Bo~Bai,
	\newblock ``Asymmetric gained deep image compression with continuous rate
	adaptation,''
	\newblock in {\em Proc. {IEEE}/{CVF} Conference on Computer Vision and Pattern
		Recognition ({CVPR})}, June 2021.
	
	\bibitem{SullivanOH2012_OverviewHighEfficiency}
	Gary~J. Sullivan, Jens-Rainer Ohm, Woo-Jin Han, and Thomas Wiegand,
	\newblock ``Overview of the high efficiency video coding ({HEVC}) standard,''
	\newblock {\em IEEE Transactions on Circuits and Systems for Video Technology},
	vol. 22, no. 12, pp. 1649--1668, Dec 2012.
	
	\bibitem{HanLM2021_TechnicalOverviewAV1}
	Jingning Han, Bohan Li, Debargha Mukherjee, Ching-Han Chiang, Adrian Grange,
	Cheng Chen, Hui Su, Sarah Parker, Sai Deng, Urvang Joshi, Yue Chen, Yunqing
	Wang, Paul Wilkins, Yaowu Xu, and James Bankoski,
	\newblock ``A technical overview of {AV}1,''
	\newblock {\em Proceedings of the {IEEE}}, vol. 109, no. 9, pp. 1--28, 2021.
	
	\bibitem{BrandFK2021_RateDistortionOptimized}
	Fabian Brand, Kristian Fischer, and Andre Kaup,
	\newblock ``Rate-distortion optimized learning-based image compression using an
	adaptive hierachical autoencoder with conditional hyperprior,''
	\newblock in {\em Proc. IEEE/CVF Conference on Computer Vision and Pattern
		Recognition Workshops (CVPRW)}, June 2021.
	
	\bibitem{BrandFK2022_LearningTrueRate_DCC}
	Fabian Brand, Kristian Fischer, Alexander Kopte, and Andre Kaup,
	\newblock ``Learning true rate-distortion-optimization for end-to-end image
	compression,''
	\newblock in {\em Proc. Data Compression Conference ({DCC})}, Mar. 2022.
	
	\bibitem{BrandFK2022_RDONetRateDistortion}
	Fabian Brand, Kristian Fischer, Alexander Kopte, Marc Windsheimer, and Andr\'e
	Kaup,
	\newblock ``{RDONet}: Rate-distortion optimized learned image compression with
	variable depth,''
	\newblock in {\em Proc. IEEE/CVF Conference on Computer Vision and Pattern
		Recognition Workshops (CVPRW)}, June 2022, pp. 1759--1763.
	
	\bibitem{RozendaalHC2021_OverfittingFunProfit}
	Ties van Rozendaal, Iris A.~M. Huijben, and Taco Cohen,
	\newblock ``Overfitting for fun and profit: Instance-adaptive data
	compression,''
	\newblock in {\em Proc. 9th International Conference on Learning
		Representations, {ICLR} 2021}, May 2021.
	
	\bibitem{SchaferPP2021_RateDistortionOptimization}
	Michael Schäfer, Sophie Pientka, Jonathan Pfaff, Heiko Schwarz, Detlev Marpe,
	and Thomas Wiegand,
	\newblock ``Rate-distortion-optimization for deep image compression,''
	\newblock in {\em Proc. IEEE International Conference on Image Processing
		(ICIP)}, Sept. 2021.
	
	\bibitem{BrandSK2022_PFrameCoding}
	Fabian Brand, Jürgen Seiler, and André Kaup,
	\newblock ``P-frame coding with generalized difference: A novel conditional
	coding approach,''
	\newblock in {\em Proc. IEEE International Conference on Image Processing
		(ICIP)}, 2022.
	
	\bibitem{ShiCH2016_RealTimeSingle}
	Wenzhe Shi, Jose Caballero, Ferenc Huszar, Johannes Totz, Andrew~P. Aitken, Rob
	Bishop, Daniel Rueckert, and Zehan Wang,
	\newblock ``Real-time single image and video super-resolution using an
	efficient sub-pixel convolutional neural network,''
	\newblock in {\em Proc. {IEEE} Conference on Computer Vision and Pattern
		Recognition ({CVPR})}, jun 2016.
	
	\bibitem{KingmaB2015_AdamMethodStochastic}
	Diederik~P. Kingma and Jimmy Ba,
	\newblock ``Adam: A method for stochastic optimization,''
	\newblock in {\em Proc. International Conference on Learning Representations
		(ICLR)}, May 2015, pp. 1--15.
	
	\bibitem{WangSB_Multiscalestructuralsimilarity}
	Z.~Wang, E.P. Simoncelli, and A.C. Bovik,
	\newblock ``Multiscale structural similarity for image quality assessment,''
	\newblock in {\em Proc. Asilomar Conference on Signals, Systems, and
		Computers}, Nov. 2003.
	
	\bibitem{AsuniG2013_TESTIMAGESLargeData}
	Nicola Asuni and Andrea Giachetti,
	\newblock ``{TESTIMAGES}: A large data archive for display and algorithm
	testing,''
	\newblock {\em Journal of Graphics Tools}, vol. 17, no. 4, pp. 113--125, 2013.
	
	\bibitem{Bjontegaard2001_CalculationaveragePSNR}
	G.~Bj{\o}ntegaard,
	\newblock ``Calculation of average {PSNR} differences between {RD}-curves,
	{VCEG-M}33,''
	\newblock {\em 13th Meeting of the Video Coding Experts Group ({VCEG})}, pp.
	1--5, Jan 2001.
	
\end{thebibliography}

\end{document}